\documentclass[twocolumn,english,aps,prl]{revtex4}
\usepackage[T1]{fontenc}
\usepackage{amsmath,graphicx,amssymb,epsfig,babel,dsfont,color}
\usepackage{textcomp}
\usepackage{epstopdf}

\begin{document}

\title{Giant near-field mediated heat flux at the nanometer scale}

\author{Konstantin Kloppstech$^1$, Nils K\"{o}nne$^1$, Svend-Age Biehs$^1$, Alejandro W. Rodriguez$^2$, Ludwig Worbes$^1$, David Hellmann$^1$}
\author{Achim Kittel$^1$}\email{kittel@uni-oldenburg.de}

\affiliation{$^1$ Institute of Physics, Carl von Ossietzky University of Oldenburg, D-26111 Oldenburg, Germany. \\
$^2$ Department of Electrical Engineering, Princeton University, Princeton, New Jersey 08544, USA.}

\date{\today}

%\pacs{44.40.+a, 65.80.-g, 68.37.Ef, 68.37.Uv}

\maketitle

%%%%%%%%%%%%%%%%%%%%%%%%%%%%%%%%%%%%%%%%%%%%%%%%%%%%%%%%%%%%%%%%%%%%%%%%%%%%%%%%%%%%
%
% Introduction 
%
%%%%%%%%%%%%%%%%%%%%%%%%%%%%%%%%%%%%%%%%%%%%%%%%%%%%%%%%%%%%%%%%%%%%%%%%%%%%%%%%%%%%
{\bf In this Letter, we report on quantitative measurements of the absolute near-field mediated heat flux between a gold coated near-field scanning thermal microscope (NSThM) tip and a planar gold sample at nanometer distances of 0.2~nm-7~nm. We find an extraordinary large heat flux which is more than five orders of magnitude larger than black-body radiation and four orders of magnitude larger than the values predicted by conventional theory of fluctuational electrodynamics. Additionally, we compare our data with different theories of phonon tunneling~\cite{PrunnilaEtAl2010,Mahan2011,BudaevBogy2011,SellanEtAl2012,ChiloyanEtAl2015} which might explain a drastically increased heat flux, but are found not to be able to reproduce the distance dependence observed in our experiment. The findings demand modified or even new models of heat transfer across vacuum gaps at nanometer distances.}

The radiative heat flux between two massive bodies held at different temperatures increases drastically when the distance $d$ between them becomes smaller than the dominant thermal wavelength $\lambda_{\rm th}$, roughly  10~\textmu m at room temperature. Consequently, the heat flux can be enhanced by many orders of magnitude compared to the heat transfer exchanged between two black bodies coupled through the far-field. This {\itshape super-Planckian} effect can be attributed to the additional contribution of evanescent waves such as frustrated total internal reflection modes, surface phonon polaritons, or hyperbolic modes~\cite{BiehsEtAl2010,BiehsEtAl2012}. The heat flux enhancement in the near-field regime has been verified by a number of recent experiments~\cite{HuEtAl2008,ShenEtAl2008,NatureEmmanuel,Ottens2011,Kralik2012,ShenEtAl2012,SongEtAl2015}. So far, the measured data has enjoyed good agreement with theoretical models of macroscopic heat transfer~\cite{HuEtAl2008,ShenEtAl2008,NatureEmmanuel,Ottens2011,Kralik2012,ShenEtAl2012,SongEtAl2015}, suggesting that super-Planckian radiation is a well-understood phenomenon and shifting the focus of current experiments/theory to the design of practical and more efficient near-field transmitters.

However, commonly used theoretical models of heat transfer are based on Rytov's theory of macroscopic fluctuational electrodynamics~\cite{Rytovbook}, which cannot fully describe heat exchange at distances down to a few nanometers. In particular, such a theory does not account for the crossover from near field to contact, in which case the objects are separated by atomic distances and heat flux is mediated by conductive transfer.
Several recent theoretical works have studied this crossover by including effects like tunneling of acoustic phonons~\cite{PrunnilaEtAl2010,Mahan2011,BudaevBogy2011,SellanEtAl2012,ChiloyanEtAl2015} and quantum effects due to the overlap of the electronic wave functions~\cite{XiongEtAl2014}, showing that the radiative heat flux can be further enhanced by orders of magnitude at distances of a few nanometers or even on the sub-nanometer level. The abovementioned experiments have confirmed the predictions of the conventional macroscopic theory~\cite{HuEtAl2008,ShenEtAl2008,NatureEmmanuel,Ottens2011,Kralik2012,ShenEtAl2012,SongEtAl2015} as they probe the near-field at much larger distances. Up to now, only one indirect measurement conducted by Altfeder {\itshape et al.}~\cite{AltfederEtAl2010} is allegedly backing the theoretical considerations on phonon tunneling using inelastic scanning tunneling microscopy. 
In contrast, compared to previous experiments our setup is the only one which can directly probe heat fluxes for distances, precisely at the interface of radiative and conductive transport.

%%%%%%%%%%%%%%%%%%%%%%%%%%%%%%%%%%%%%%%%%%%%%%%%%%%%%%%%%%%%%%%%%%%%%%%%%%%%%%%%%%%%
%
% Letter Summary
%
%%%%%%%%%%%%%%%%%%%%%%%%%%%%%%%%%%%%%%%%%%%%%%%%%%%%%%%%%%%%%%%%%%%%%%%%%%%%%%%%%%%%
\begin{figure}[Hht]
	\includegraphics[width=0.45\textwidth]{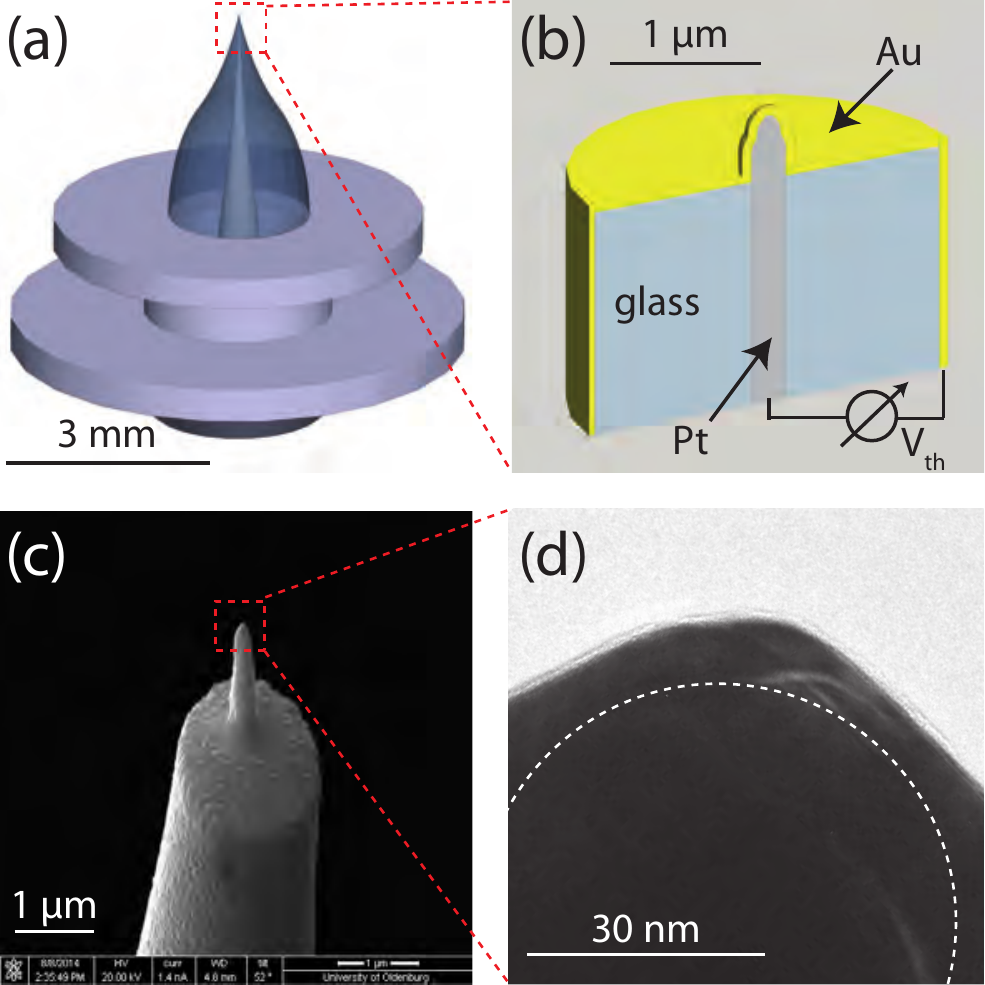}
	\caption{{\bf NSThM-probe.} (a) Sketch of the probe in its Omicron-type tip holder. (b) Schematic cross-section of the sensing end of the probe. A thermocouple is formed where the gold coating separates from the platinum core. (c) SEM micrograph of a typical NSThM probe. (d) TEM image --- more precisely, the shadow because the tip is too thick to be electron transparent --- of the tip of a typical NSThM probe indicating a radius of curvature of about 30~nm (dashed semicircle). Here the axis of rotational symmetry lies in the vertical direction. }
	\label{Fig:Probe}
\end{figure}

%%%%%%%%%%%%%%%%%%%%%%%%%%%%%%%%%%%%%%%%%%%%%%%%%%%%%%%%%%%%%%%%%%%%%%%%%%%%%%%%%%%%
%
% Experimental Setup
%
%%%%%%%%%%%%%%%%%%%%%%%%%%%%%%%%%%%%%%%%%%%%%%%%%%%%%%%%%%%%%%%%%%%%%%%%%%%%%%%%%%%%
Our experiment is performed with a custom-built NSThM under highly controlled conditions in ultra-high vacuum with 
a typical working pressure of $10^{-10}$ mbar. The setup is based on a commercial scanning tunneling microscope (STM).
As depicted in Fig.~\ref{Fig:Probe} (a) and (b), the home-made STM probes consist of a platinum wire, molten into a glass capillary, pulled sharp with a pipette puller and are
then coated with $100~{\rm nm}$ of Au by means of e-beam evaporation \textit{ex situ}. At the point where the Au film separates from the Pt-core, a thermocouple is formed. This probe design allows for local heat flux measurements in addition to its STM ability~\cite{UliEtAl2008,KittelEtAl2008}. The heat flux coupled into the tip apex drains towards the back side of the tip holder causing a temperature difference between them which, finally, is generating a thermovoltage $V_{\rm th}$. A scanning electron microscope (SEM) image of such a probe is depicted in Fig.~\ref{Fig:Probe} (c). The protruding part of the probe is typically about 1 -- 2~\textmu m in length and 300 -- 700~nm in diameter (at the base). The radius $r$ of the tip apex is typically about $30~{\rm nm}$~\cite{WorbesEtAl2013}, as shown in the transmission electron microscope (TEM) micrograph in Fig.~\ref{Fig:Probe} (d).

Our probes are able to detect heat fluxes down to 4~nW and heat conductances down to $24~{\rm pW/K}$ at $50~{\rm Hz}$ bandwidth. As we will see below, this sensitivity of the probe is not sufficient to measure radiative heat fluxes predicted by fluctuational electrodynamics. Concerning the heat fluxes, we achieve a lateral resolution of $6~{\rm nm}$ when a temperature difference $\Delta T$ between probe and sample is applied~\cite{KittelEtAl2008,UliEtAl2008}. The topographic information can be measured at the same time using the STM ability of our probe which features atomic resolution.

%%%%%%%%%%%%%%%%%%%%%%%%%%%%%%%%%%%%%%%%%%%%%%%%%%%%%%%%%%%%%%%%%%%%%%%%%%%%%%%%%%%%
%
% Measurements 
%
%%%%%%%%%%%%%%%%%%%%%%%%%%%%%%%%%%%%%%%%%%%%%%%%%%%%%%%%%%%%%%%%%%%%%%%%%%%%%%%%%%%%

\begin{figure}[Hhbt]
	\centering
	\includegraphics[width=0.45\textwidth]{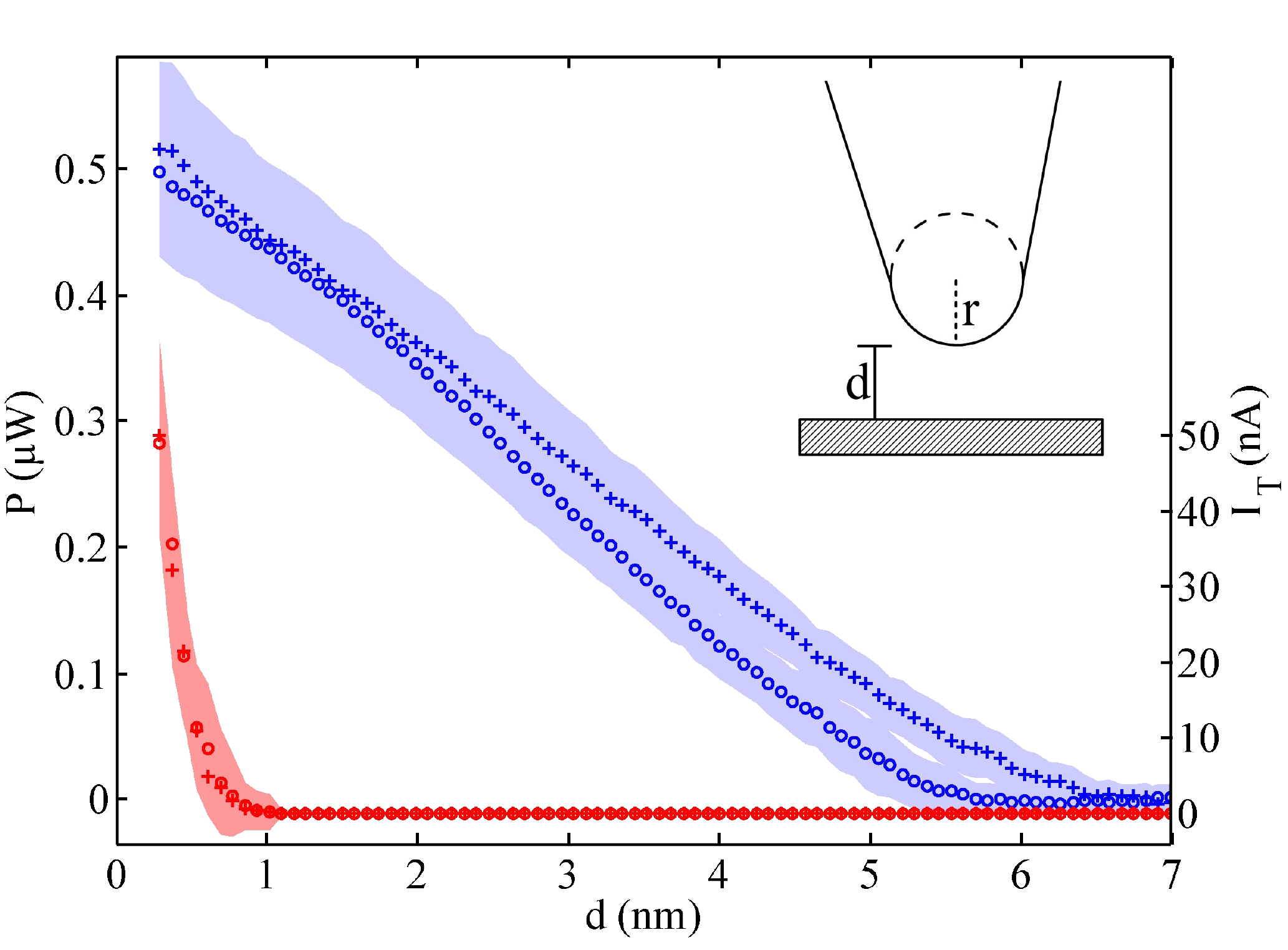}
	\caption{{\bf Gap-dependent heat flux and tunneling current.} Measured average heat flux power $P$ (upper curves with respect to the axis on the left-hand side) 
         and tunneling  current $I_{\rm T}$ (lower curves with respect to the axis on the right-hand side) as a function of distance $d$ for approaching (circles) and 
         retracting (crosses) direction together. The sample given by a 200~nm gold-film on a mica substrate is cooled down to 120~K,
         whereas the temperature of the probe is held at ambient temperature so that $\Delta T = 160\,{\rm K}$. The shaded areas quantify the uncertainties:
         In case of the tunneling current the uncertainty is given by its standard deviation, whereas the relative error of the heat flux measurement 
         is calculated via Gaussian error calculus for each distance step. The certainty of the value for the distance $d = 0$~nm is limited by the certainty 
         of the value for the work function for gold~\cite{Behm1990}. From which we estimate a relative error in $d = 0$~nm of $\Delta d = 90~{\rm pm}$. 
         Inset: Sketch of the probe and the sample.
}
\label{Fig:thermovoltage}
\end{figure}

The measured change of the probe-sample heat current $\Delta P$ in the 
distance regime of 0.2 - 7~nm is approximately 0.5~\textmu W as shown 
in Fig.~\ref{Fig:thermovoltage}. This corresponds to a heat transfer coefficient $h_{\rm nf}$ through the vacuum gap by near-field interactions of
\begin{equation}
 h_{\rm nf} = \frac{\Delta P}{A \Delta T} = 1.11\times10^6~\frac{\rm W}{{\rm m}^2 {\rm K}}
\end{equation}
when using $\Delta P = 0.5~$\textmu W and assuming a disk-shaped effective heat flux area $A$ of the tip with $r = 30~{\rm nm}$ and a temperature difference of $\Delta T = 160~{\rm K}$, since $T_{\rm probe} = 280~{\rm K}$ and $T_{\rm sample} = 120~{\rm K}$. In contrast, the heat transfer coefficient between two black bodies at the same temperatures can be estimated to be
\begin{equation}
 h_{\rm BB} = \sigma_{\rm BB} \frac{T^4_{\rm probe} - T^4_{\rm sample}}{\Delta T} = 2.10~\frac{\rm W}{{\rm m}^2 {\rm K}}
\end{equation}
using the Stefan-Boltzmann constant $\sigma_{\rm BB} = 5.67\times10^{-8} {\rm W}/{\rm m^2} {\rm K}^4$. 
Hence, the measured heat transfer coefficient of the vacuum gap is
about $5\times10^5$ times larger than the black-body value. Thus, our NSThM technique yields by far the largest heat flux level compared to other near-field experiments~\cite{HuEtAl2008,ShenEtAl2008,NatureEmmanuel,Ottens2011,Kralik2012,ShenEtAl2012,SongEtAl2015}. So far, these have measured heat fluxes up to approximately $100$ times the black-body value~\cite{Kralik2012}, albeit at much larger distances. As we show below, this value is four orders of magnitude larger than that obtained using conventional macroscopic fluctuational electrodynamics. However, theoretical models based on phonon tunneling can predict such large values. Furthermore, we find at close distances up to $d = 2~\rm nm$ an almost linear decay of the heat flux, which means that we can exclude algebraic decays of the form $d^{-n}$ with $n \geq 1$, but we cannot exclude exponential decays.

We want to emphasize that the measured heat transfer cannot be caused by Joule heating from tunneling electrons. The maximum power of Joule heating by the tunneling electrons can be estimated to be $P_{\rm e^{-}} = V I_{\rm T} = 30\,{\rm nW}$ ($V = 600\,{\rm mV}$ and $I_{\rm T} = 50\,{\rm nA}$) which is about six percent of the maximum heat flux. 
Furthermore, Fig.~\ref{Fig:thermovoltage} shows the typically observed exponential decay of the tunneling current $I_{\rm T}$ at distances below $d = 1\,{\rm nm}$. At larger distances no current is detectable anymore (below 0.5~pA). Thus, the massive heat flux and its distance dependence cannot be explained by the exchange of electrons.

%%%%%%%%%%%%%%%%%%%%%%%%%%%%%%%%%%%%%%%%%%%%%%%%%%%%%%%%%%%%%%%%%%%%%%%%%%%%%%%%%%%%
%
% Diskussion 
%
%%%%%%%%%%%%%%%%%%%%%%%%%%%%%%%%%%%%%%%%%%%%%%%%%%%%%%%%%%%%%%%%%%%%%%%%%%%%%%%%%%%%

\begin{figure}[Hhbt]
  \begin{center}
  \includegraphics[width=0.5\textwidth]{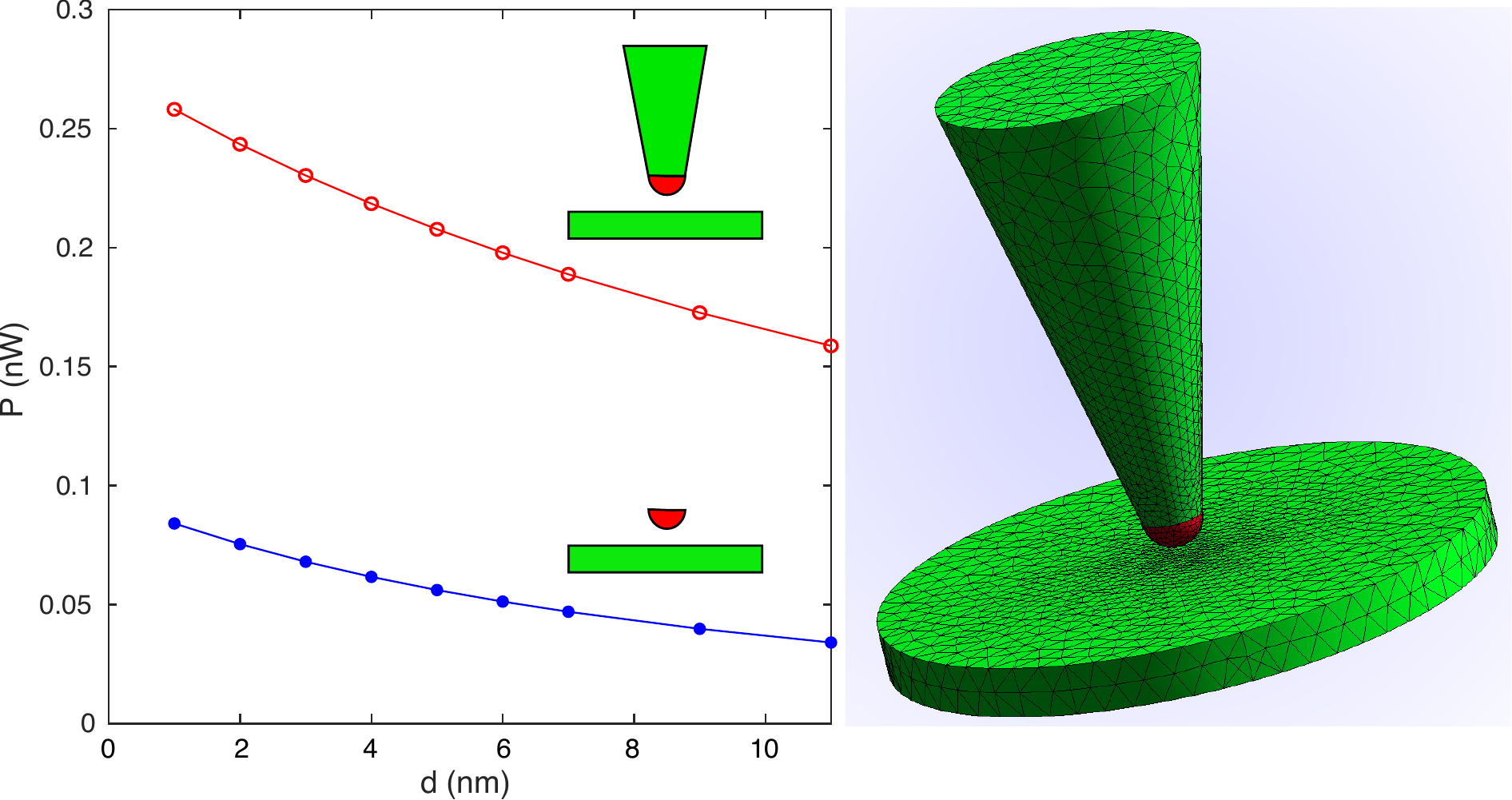}
  \end{center}
  \caption{{\bf Theoretical results of the transfered heat flux.} Sketch of the considered geometry (righthand side) and numerical results 
           using exact numerical calculations for the spherical tip and the cone-like protruding part. 
           We have dyed the different parts of the probe (see inset) with the same colors as used for the lines in plots. 
           The parameters of the tip are the following: the foremost part is modeled by a sphere of radius of 30~nm, the protruding conical 
           part has a length of 300~nm with a radius at the base of 87.5~nm. \label{Fig:Modeling}}
\end{figure}

Now, we want to compare commonly discussed theoretical models with our experimental data. Let us first stick to the conventional macroscopic theory: In Fig.~\ref{Fig:Modeling} we show exact numerical for the radiative heat flux using a boundary element method in order to model the geometry of our probe~\cite{RodriguezEtAl2012b,RodriguezEtAl2012b}. Obviously, the values for the heat flux are between 0.2~nW($d = 5\,{\rm nm}$) and 0.25~nW($d = 0.5\,{\rm nm}$) which corresponds to heat transfer coefficients $h_{\rm nf}$ of about 440-550~${\rm W}/{\rm m}^2 {\rm K}$. Therefore, the change of the probe-sample heat current as predicted by fluctuational electrodynamics is only 110~${\rm W}/{\rm m}^2 {\rm K}$, whereas the measured value is $1.1\,\times10^6~{\rm W}/{\rm m}^2 {\rm K}$. It is interesting that the distance dependence of the heat flux found Fig.~\ref{Fig:Modeling} is similar to the measured distance dependence, but the heat flux level is obviously four orders of magnitude too small. Thus, the curvature of the tip cannot account for the discrepancy between the experimental data and the theory.

It might be argued that nonlocal response of the permittivity should be taken into account~\cite{Volokitin2001,KittelEtAl2005,Chapuis2008}. 
Since nonlocal effects cannot be included easily in the exact numerical scheme, we have also modeled our sensor using the  so-called proximity approximation (PA),  which can be applied in the near-field regime when the distance between two objects is much smaller than their curvature~\cite{Krueger2011,OteyFan2011,BiehsRoughness2010a}.  Note, that this approximation has succesfully been applied in all experiments using a spherical probe~\cite{ShenEtAl2008,NatureEmmanuel,ShenEtAl2012,SongEtAl2015}.  In our case, however, we obtain results for the whole probe which show a similar distance dependence as the exact results in Fig.~\ref{Fig:Modeling}, but which overestimate the contribution of the tip and lead to errors on the order of 400\% (see Suppl.~Mat.). Similar deviations of PA and exact results were already observed in a sphere-plate geometry (see Ref.~\cite{OteyFan2011}) and here can be traced back to the fact that the foremost part of the tip is extremely small compared with $\lambda_{\rm th}$ so that it acts more like a dipole than a macroscopic sphere. Nonetheless, one can still use the PA to estimate the heat flux in our setup, keeping in mind that it overestimates the heat flux level. Now, within this approximation we have included nonlocal effects using the Lindhard-Mermin model~\cite{Chapuis2008}  (see Suppl.~Mat.), and we find that while these nonlocal effects increase the heat flux, as expected~\cite{Volokitin2001,Chapuis2008}, they turn out to be relatively weak for the considered distances. Hence, we find that the conventional macroscopic model of heat transfer greatly underestimates the heat flux found in our experiments. We note and emphasize that the above calculations in Fig.~\ref{Fig:Modeling} make no approximation and fully account for flux mediated by surface plasmon polaritons, though these tend to be negligible for gold at room temperature~\cite{Chapuis2008}. In what follows, we consider a number of currently accepted models of phonon (conductive) transfer and argue that they too cannot explain the above mentioned enhancement:

(A) Prunnila and Meltaus~\cite{PrunnilaEtAl2010} have studied the tunneling of acoustic phonons between piezoelectric materials. They report an approximate $1/d^3$ distance dependence. Making the same estimation for the effective tip area $A$ as in (i) we find
that for piezoelectric materials studied in~\cite{PrunnilaEtAl2010} a power transfer of about 13~nW could be expected for $r = 30~{\rm nm}$ at $d = 1~{\rm nm}$ ($h_{\rm nf}=2.9\times10^4~{\rm W}/{\rm m}^2 {\rm K}$). This value is too small to explain our data and the distance dependence does not agree with the measured one. But this is not surprising, since in our experiment we are not using piezoelectric materials but metals, meaning that this theory cannot be applied directly. 

(B) Another approach is given by Sellan {\itshape et al.}~\cite{SellanEtAl2012} who consider the phonon tunneling between two silicon half spaces through a vacuum gap using lattice dynamics calculations. The authors report a heat flux by phonon tunneling (resulting in a heat transfer coefficient $h_{\rm nf}$ of $5.3\times10^8~{\rm W}/{\rm m}^2~{\rm K}$) which is five orders of magnitude larger than the conventional radiative heat flux at $d = 0.1~{\rm nm}$. Although the reported enhancement is about two orders of magnitude larger than ours, this effect is only observable at distances smaller than 0.2nm, above which the theory is well described by macroscopic fluctuational electrodynamics. Hence, phonon tunneling within this model also cannot explain our enhancement which occurs at distances up to 5~nm. Furthermore, the calculations were only done for Si using a specific Stillinger-Weber potential usually used in bulk material. It should be mentioned that using another atomistic simulation method, it was shown very recently that for polar materials like SiC phonon tunneling only slightly increases the heat flux with respect to Rytov's theory in the distance regime between 0.2~nm and 1~nm~\cite{ChiloyanEtAl2015}.

(C) The approach of Mahan~\cite{Mahan2011} based on image potentials considers heat tunneling between a metal and alkali halides so that it is again not directly applicable to our experiment. The theory predicts a $1/d$ dependence with very large heat fluxes by phonon tunneling 
even for several nanometers. At distances of 2--3~nm, the heat flux drops by a factor of 10 compared to the boundary Kapitza conductance between metals and alkali halide, which yields a heat-transfer coefficient on the order of  $h_{\rm nf}=10$ -- $100\times10^6~{\rm W}/{\rm m}^2 {\rm K}$. Hence, while such a model does predict heat fluxes of similar magnitudes as those obtained here at certain separations, we find that it cannot explain the distance dependence observed in our measurements. Furthermore, this model cannot be directly applied to the case of two metal surfaces.

(D) Yet another but very general description for phonon tunneling was proposed by Budaev and Bogy~\cite{BudaevBogy2011} 
%(a similar approach was recently published in Ref.~\cite{EzzahriJoulain2014}) 
based on a very elementary classical oscillator model for describing lattice vibrations. The authors find in their model that the heat flux by phonon tunneling scales like $1/d^8$. For Si they find that the heat flux by phonon tunneling equals the conventional radiative heat flux at $d = 5~{\rm nm}$ ($h_{\rm nf}\approx10^5~{\rm W}/{\rm m}^2 {\rm K}$) and dominates the heat flux for smaller distances. This model obviously already gives different results than predicted by the much more elaborate method (B). Furthermore, the predicted $1/d^8$ distance dependence does not agree with our data. 

Summarized, we see that the theoretical models proposed in the literature so far cannot be directly applied and therefore cannot describe the data found in our present experiment. We believe that this, along with previous results on a related experiment~\cite{WorbesEtAl2013}, provide the basis and motivation for further theoretical exploration of heat transfer mechanisms in this crossover regime where both radiative and conductive effects can coexist, and can be greatly affected by geometry. 

%%%%%%%%%%%%%%%%%%%%%%%%%%%%%%%%%%%%%%%%%%%%%%%%%%%%%%%%%%%%%%%%%%%%%%%%%%%%%%%%%%%%
%
% Conclusion
%
%%%%%%%%%%%%%%%%%%%%%%%%%%%%%%%%%%%%%%%%%%%%%%%%%%%%%%%%%%%%%%%%%%%%%%%%%%%%%%%%%%%%

In conclusion, we have measured the near-field mediated heat flux between a gold coated near-field scanning microscope tip and a gold sample at distances of a few nanometers. The measured values for the heat flux are four orders of magnitude larger than predicted by fluctuational electrodynamics and five orders of magnitude larger than the black-body value.  A comparison with current models of phonon tunneling shows that they can explain why the heat flux can be much larger than predicted by fluctuational electrodynamics, but these models typically predict an algebraic decay of the heat flux which is in contradiction to the measured decay. Given the current lack of appropriate theoretical models that can span this range of distances and geometries, the question of whether the measured heat flux can be explained by phonon or photon tunneling, or whether there is yet another unknown mechanism at play, remains open. We hope, however, that this work will serve to motivate further study of this increasingly important yet unexplored regime.

%%%%%%%%%%%%%%%%%%%%%%%%%%%%%%%%%%%%%%%%%%%%%%%%%%%%%%%%%%%%%%%%%%%%%%%%%%%%%%%%%%%%
%
% Methods
%
%%%%%%%%%%%%%%%%%%%%%%%%%%%%%%%%%%%%%%%%%%%%%%%%%%%%%%%%%%%%%%%%%%%%%%%%%%%%%%%%%%%%

\section{Methods}

We use 1w measuring techniques along with an adapted hot-wire method.

{\bf Characterization and calibration of the probe.} In a first step our probe is characterized \textit{in situ} just before the actual measurement to obtain the ratio $\varepsilon$ between the heat flux through the probe and the hereby generated thermovoltage $V_{\rm th}$. To this end, we use $1\omega$ measuring techniques along with an adapted hot-wire method. The details of this calibration method can be found in Ref.~\cite{Kloppstech}. For the NSThM probe used in our measurements we find a calibration factor of $\varepsilon = 0.43 $~\textmu W/\textmu V. The heat flux through the probe is given by $P = \varepsilon V_{\rm th}$. The purely near-field contribution to the heat transfer is detected by subtracting the heat flux at larger distances (typically a few tens of nanometer) realizing that the near-field effect has fallen below the detection limit at these distances. Our calibration method enables us to measure absolute heat fluxes between the probe and the sample with a relative uncertainty of about 14\% (details of the error analysis leading to this uncertainty can be found in Ref.~\cite{Kloppstech}). 

{\bf Characterization of the sample.} The sample used in our measurement consists of a 200~nm Au layer deposited \textit{ex-situ} via e-beam evaporation on a cleaved and heated mica substrate, leading to a monocrystalline Au(111) surface. After cleaning (sputtering with Ar ions and annealing) it shows wide atomically flat areas and the common $22\times\sqrt{3}$ surface reconstruction. In a next step, the sample is cooled down to $120~{\rm K}$, while the probe is held close to ambient temperature leading to a probe-sample temperature difference $\Delta T =160~{\rm K}$. 

{\bf Measurement of the heat flux.} Using the STM ability of our probe, first an atomic flat area of about $75\times75~{\rm nm}^2$ is localized where we then perform our measurements of the heat flux from the probe tip to the sample. This is done, firstly, by lifting the probe $7~{\rm nm}$ from the tunneling distance (tunneling current $I_{\rm T}=1$~nA, bias voltage $V_{\rm T}=600$~mV). Then, the sample is approached stepwise at a maximum slew rate of $90~{\rm nm/s}$ until a threshold of the tunneling current of $50~{\rm nA}$ is reached which corresponds to a sample-probe distance $d$ of about $0.2~{\rm nm}$. While approaching the surface the thermovoltage is acquired after a settling time with an integration time of 20~ms, both, using distance steps of about 0.08~nm. Then the probe is retracted again and the heat flux is measured using the same measurement procedure. Finally, the tip is brought back to its original tunneling distance. Averaged data of the heat flow for approaching and retracting measurements are shown in Fig.~\ref{Fig:thermovoltage}. Note, that only those measurements are taken into account where the difference in the piezo stroke at the beginning and at the end of the measuring cycle is less than $50~{\rm pm}$. This is done in order to avoid drift artefacts, spoiling the averaging procedure of 100 measurements for each direction.
We emphasize that here $d = 0$~nm corresponds to the inter-atomic distance in bulk gold. Therefore, the value $d = 0$~nm denotes the distance where the electrical conductance of a single Au atom ($I_{\rm T}=$46.5~\textmu A; 600~mV) is reached in the STM.

{\bf Numerical modelling.}
 The theoretical predictions of heat flux in Fig.~\ref{Fig:Modeling} were obtained via a recently developed fluctuating--surface current formulation of heat transfer that is applicable to arbitrary geometries and has been validated against well-established results in more conventional geometries, including planar and spherical objects~\cite{RodriguezEtAl2012,RodriguezEtAl2012b}. Our results were obtained using a free numerical implementation of this method based on the boundary-element method, which discretizes the object surfaces (the scattering unknowns) using localized Rao-Wilton-Glisson (RWG) basis functions. The model geometry follows the same parameters as the experiment, with material parameters described via the standard Drude model of Au, except that the cone height is capped at 300~nm and the planar sample is taken to be a cylinder with a finite radius of 600~nm and a thickness of 50~nm. We find that larger cone heights and sample diameters have a negligible impact on the overall heat transfer, as does the contribution of the planar base (see Suppl.~Mat.). Fig.~\ref{Fig:Modeling} shows a schematic of the discretized geometry, which is chosen to yield converged results with respect to mesh size (resolution) and geometric parameters. Comparison between the numerical and PA predictions shows that the latter greatly overestimates the impact of the sphere-tip in comparison with the conical section as well as the overall flux (see Suppl.~Mat.).

%%%%%%%%%%%%%%%%%%%%%%%%%%%%%%%%%%%%%%%%%%%%%%%%%%%%%%%%%%%%%%%%%%%%%%%%%%%%%%%%%%%%
%
% Acknowledgements
%
%%%%%%%%%%%%%%%%%%%%%%%%%%%%%%%%%%%%%%%%%%%%%%%%%%%%%%%%%%%%%%%%%%%%%%%%%%%%%%%%%%%%

\section{acknowledgments}
We thank Martin Holthaus and Philippe Ben-Abdallah for valuable discussions, Holger Koch for technical support, and J\"urgen Parisi for constant support. This work was supported by the DFG (Deutsche Forschungsgemeinschaft) (Project No. KI 438/8-1) and the EWE AG Oldenburg/Germany (EWE-Nachwuchsgruppe). AWR was supported by the National Science Foundation under Grant No. DMR-145483.

%%%%%%%%%%%%%%%%%%%%%%%%%%%%%%%%%%%%%%%%%%%%%%%%%%%%%%%%%%%%%%%%%%%%%%%%%%%%%%%%%%%%
%
% Authors contribution
%
%%%%%%%%%%%%%%%%%%%%%%%%%%%%%%%%%%%%%%%%%%%%%%%%%%%%%%%%%%%%%%%%%%%%%%%%%%%%%%%%%%%%

\section{Authors contributions}

The actual experiment were performed and analyzed by K.K. and N.K. under supervision of A.K.
L.W. and D.H. have developed the different experimental procedures and assisted with analysis and discussion of the results.
Modelling was done by A.W.R. and S.-A.B. The manuscript was written by S.-A.B., K.K., and A.K. All
authors discussed the experimental results/implications and commented on the manuscript at all stages.

%%%%%%%%%%%%%%%%%%%%%%%%%%%%%%%%%%%%%%%%%%%%%%%%%%%%%%%%%%%%%%%%%%%%%%%%%%%%%%%%%%%%
%
% Additional Information
%
%%%%%%%%%%%%%%%%%%%%%%%%%%%%%%%%%%%%%%%%%%%%%%%%%%%%%%%%%%%%%%%%%%%%%%%%%%%%%%%%%%%%

\section{Additional information}

Supplementary information is available in the online version of the paper. Reprints
and permissions informatino is available online at www.nature.com/reprints. Correspondence
and requests for materials should be addressed to S.-A.B.\ and A.K.

%%%%%%%%%%%%%%%%%%%%%%%%%%%%%%%%%%%%%%%%%%%%%%%%%%%%%%%%%%%%%%%%%%%%%%%%%%%%%%%%%%%%
%
% Financial interest
%
%%%%%%%%%%%%%%%%%%%%%%%%%%%%%%%%%%%%%%%%%%%%%%%%%%%%%%%%%%%%%%%%%%%%%%%%%%%%%%%%%%%%

\section{Competing financial interests}

The authors declare no competing financial interests.

\end{document}